# Features of formation of the Al-Cu single crystals structure under the action of pressure and gravitational field on the crystallization process.

V. O. Esin, A. S. Krivonosova, I. Zh. Sattybaev, L. V. Elokhina, T. G. Fedorova.

*Institute of Metal Physics, Ural Division of the Russian Academy of Sciences, Ekaterinburg, Russia.*
*E-mail: yesin@imp.uran.ru*

X-ray structural and spectral analyses, X-ray diffraction topography and metallography were used to study dendrite structure of Al-4 wt % Cu single crystals grown by the Bridgman technique at a rate of 1 mm/min and on seeds with <001> orientation. The single crystals were obtained upon the crystallization of the melt under the action of a pressure and the gravitational-field component directed along the surface of the crystallization front. Non-uniform distribution of copper in the solid solution and the eutectic content has been found in transverse section of single crystals and strong increase in their concentration was observed at the lateral side of the samples in the direction of the gravitational-field component action. Distribution of copper and the eutectic phase depend on pressure.

## INTRODUCTION

In our previous works [1,2] performed on aluminum single crystals with small copper additions (0.001 – 0.01 wt % Cu) the process of crystallization of the melt was experimentally studied in the presence of component of the earth gravitational field directed along the surface of the crystallization front. A non-uniform distribution of dislocations and copper solute with a strong increase in their concentration near the lateral surface of the crystal was found in the direction of the gravitational-field component action – "transverse sedimentation". It is shown that the most efficient spatial separation of the alloy components and redistribution of dislocations are observed at a certain pressure that is "optimum" under given experimental conditions (when the interface boundary exhibits an extremely high mobility). The redistribution of structural defects is increasing with the increase of gravitational-field component and is decreasing with the increase of crystallization rate.

In this connection, it is of interest to study the structure of single crystals with dendrites growth form so far as sedimentation of the solute component must to act on segregation in the melt of two-phase zone. In this work we studied the effect of low excessive pressure and gravitational-field component on the structure of single crystals of binary Al – 4 wt % Cu alloy.

## EXPERIMENTAL

The single crystals of Al -4 wt % Cu alloy 120 mm long and 20 mm in diameter were grown by the modified Bridgman technique [1] at a rare of 1 mm/min using graphite crucibles and seeds with <001> orientation, in vacuum and in argon atmosphere at pressures of 0.15 and 0.25 MPa.

The gravitational-field component directed along the surface of the crystallization front was generated by an inclination of the crystallization chamber and the heating furnace with respect to the direction of the earth gravity force $g_0$ by an angle $\varphi=20^0$ and had a value on the order of ~0.342 $g_0$.

In order to carry out structural studies on single crystals, we used metallography, X-ray diffraction topography, X-ray structural and spectral analyses. To study the structure, the samples were spark-cut along the longitudinal section of crystals in this case coincided with the plane of inclination of the crystallization chamber and the heating furnace.

## RESULTS AND DISCUSSION

*1. Particularity of the dendrite structure of single crystals.*

The analysis of microstructure shows that cellular-dendritic structure is formed in all single crystals of Al-4% Cu alloy. Such a structure is formed during the growth of dendritic ensembles [3,4] when primary dendritic arms align along the growth direction and secondary arms form a rectangular network in the cross section. In the process of solidification of the melt, interdendritic spaces remain liquid until primary dendritic arms grow in the melt over on a significant distance, depending on the width of the two-phase zone. The subsequent solidification occurs in the two-phase region, in which segregation takes place concurrently under the action of lateral diffusion flows. The segregation leads to changes in the melt composition of interdendritic spaces (from $C_0/k$ about dendritic top to the eutectic concentration in a bottom) and to the precipitation of the eutectic phase containing intermetallic $CuAl_2$.

The metallographic studies allowed us to determine some morphological trends in the formation of dendritic cells that are related to the effect of gravitation-field component on the solidification process. The presence of two field (domain) with different of the growth directions of the primary dendritic arms are distinctive feature of this structure (Fig.1). In 2/3 of the crystals volume primary dendritic arms (fIeld 1) grows in the direction [001], which coincide with a direction of the heat removal from of the crystallization front. In 1/3 of the crystals volume primary dendritic arms (field 2) grows in the direction [010] coincide with the direction of action of gravitation-field component. We suppose that this field 2 is forming at the expense of development of the concentration supercooling of a melt as a result of the transverse sedimentation of copper. It is confirmed of the experimental dates of the redistribution of copper in the single crystals.

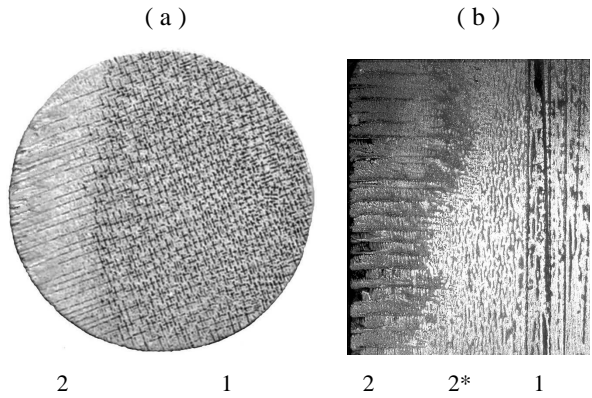

Figure 1. Macrostructure of the single crystals grown under the pressure of 0.25 MPa. (a – transverse and b – longitudinal polished sections).

Unstable subdendritic zone ($2^*$) is presents between field 1 and field 2. In this domain a supercooling of the melt in directions [001] and [010] is like. The solidification of the melt with various copper concentrations (domain 1, 2) leads to the formation of the dendritic structure of various dispersion.

*2. Transverse sedimentation of the copper.*

To study the distribution of solute component of alloy, direct measurements of the copper concentration in solid solution by the method of electron microprobe analysis and indirect measurements of the copper by measurements of the lattice parameter of the α-solid solution were carried out. The copper also is inserted at the intermetallic compound $CuAl_2$, therefore the content of the eutectic phase were measured.

The lattice parameter of the solid solution in the Al-4% Cu alloy single crystals was determined using the X-ray diffraction analysis carried out on a DRON – 3 diffractometer. In order to count of the lattice parameters into a copper concentration known dependences were used [3]. A figur 2 is showing the curves of the distribution of a copper in the transverse sections (at the distance 40 mm from the seed) of the single crystals grown at the various pressures. These curves confirm that in the process of crystal growth the redistribution of copper in its transverse section occurs in the direction of the gravitational-field component.

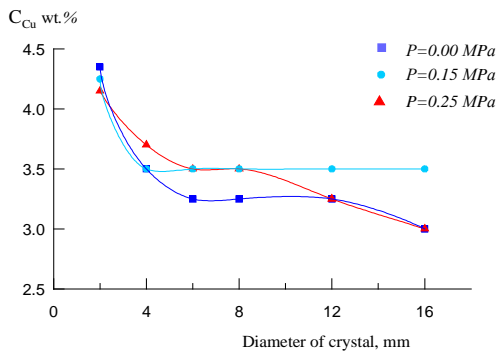

Figure 2. Distribution of concentration of copper in the solid solution of the single crystals grown under various pressures along the diameter of transverse polished section.

These results are also corroborated by the results of quantitative electron microprobe analysis. A fidisplays the curves of the distribution of a copper in the longitudinal section (at the distance 60 mm from the seed) of the single crystals grown at the various pressures.

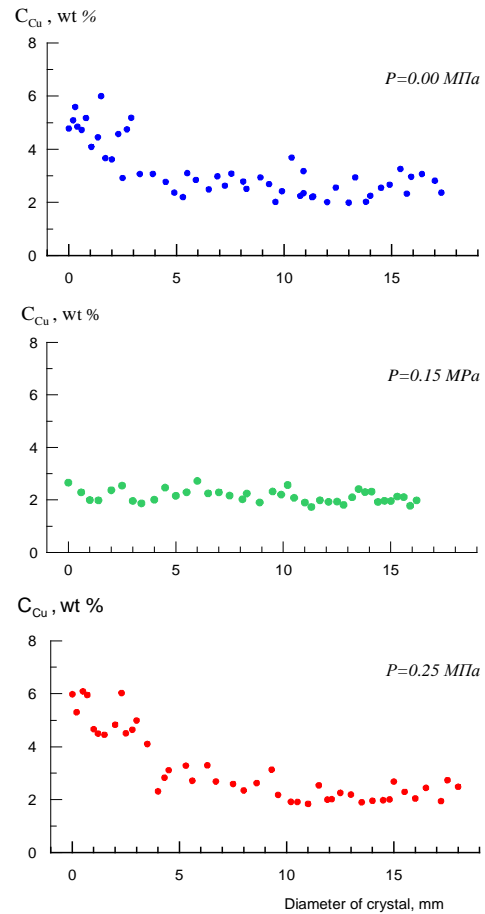

Figure 3. Distribution of concentration of copper in the solid solution of the single crystals grown under various pressures along the diameter of transverse polished section.

The measurements were performed on a SUPERPROBE–733 X-ray microanalyser along the horizontal diameter of the crystal that coincided with the direction of the action of the gravitational-field component oriented along the surface of the crystallization front. It was found the concentration of a copper markedly increase in the direction of the gravitational field.

The content of the eutectic phase was measured using known metallographic techniques on a NEOPHOT–2 optical microscopes. A figure 4 displays the distribution curves of quantity of eutectic in the longitudinal section at the distance (a)-40 and (b)-100 mm from the seed of the single crystals grown at the various pressures. The results of the investigation of the eutectic distribution show that the quantity of eutectic phase markedly increase in the direction of the gravitational field. It leads to additional increase in the general redistribution of copper in a crystals under the action of gravitational field.

Thus, the experimental data received by various methods of research are in the good consent among themselves. It allows to draw the conclusion that and at the dendritic form of growth (as well as at the flat shape of an interface between liquid and solid phases) also is observed effective separation of components of an alloy with increase in concentration of heavier element near to a lateral surface of a crystal in a direction of action of a field.

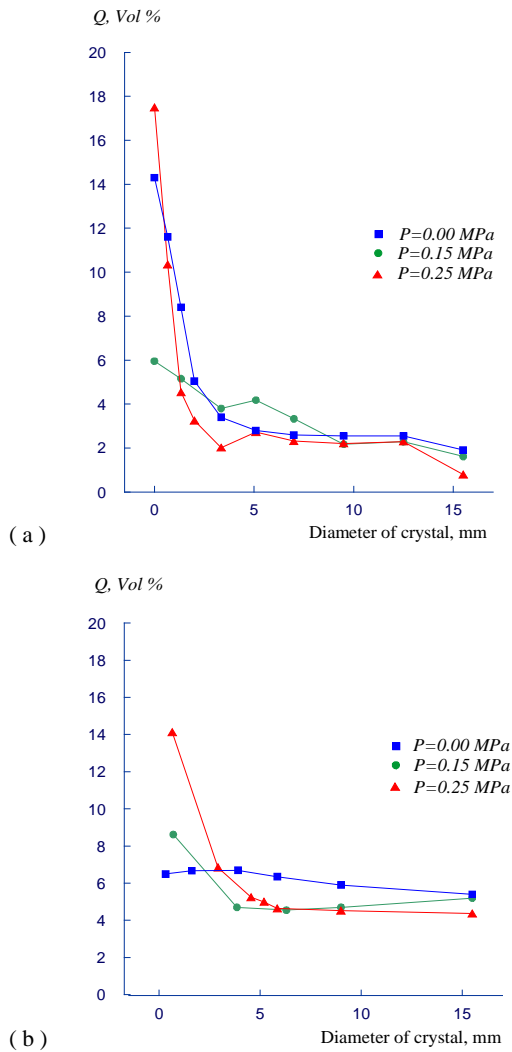

( a )

( b )

Figure 4. Distribution of eutectic phase of single crystals under various pressures along the diameter of longitudinal polished section.

*3. Effect of the pressure on the sedimentation of copper.*

As was shown earlier [1,2] that the degree of separation of the alloy components in the Al with small concentration of the copper (upon the plane shape of the crystallization front) depends on the magnitude of pressure at which the crystallization of the melt is realized and reaches a maximum at a certain pressure that is "optimum" for given conditions of the process of the phase transformation. An above-mentioned experimental results shows (fig. 2–5) was depended on the pressure.

As was shown earlier [1,2] that the degree of separation of the alloy impurities in the aluminum–copper system (upon the plane shape of the crystallization front) depends on the magnitude of pressure at which the crystallization of the melt is realized and reaches a maximum at a certain pressure that is "optimum" for given conditions of the process of phase transformation (the rate of the crystal growth and the crystallographic orientation of the direction growth) [3].

Above-mentioned experimental results (Fig. 2–4) shows that the gravitational separation in the Al – 4 wt % Cu alloy (upon the dendritic form of the crystal-melt interface boundary) also depend on the pressure. As measurements show (Fig. 4), in single crystals grown under the pressure of 0.25 MPa at most quantity of eutectic phase is formed. In single crystals grown under the pressure of 0.15 MPa lesser quantity of eutectic phase is formed. This difference is kept on all along of crystal. This is also corroborated by the results of quantitative electron microprobe analysis (Fig. 3). The copper concentration in the single crystals grown at a pressure of 0.25 MPa was found to markedly increase in the direction of the gravitational-field component and reach to limit of solubility in solid phase. While the concentration of the copper is remain of invariable.

It is known that in growing single crystals by the Bridgman technique the impurities with the distribution coefficient unity is less than unit are accumulated at the end of the sample due to the rejection by the moving crystal-melt interface boundary. In the experiments performed in the presence of gravitational-field component directed along the plane of the crystallization front, in additional to a conventional increase in the solute component of the alloy along the crystal length, a strong nonuniformity of the copper distribution was found in the direction perpendicular to the direction of growth (in the transverse section of the single crystal).

In the single crystals grown at a pressure of 0.25 MPa, the copper is concentrated in a very narrow near-surface zone with a characteristic crescent shape (Fig. 1) in the transverse section that is retained over the whole crystal. While in the single crystals obtained at a pressure of o.15 MPa redistribution of copper is pronounced appreciably weaker and is not virtually revealed in the upper half of the crystal (Fig. 4).

If the redistribution of copper in the transverse direction is indeed more efficient in the crystals grown under 0.25 MPa pressure as compared to the crystals grown in vacuum and particularly in 0.15 MPa, then in the end parts of these crystals the amount of copper accumulated should be less, since in this care their total amount should remain unaltered. Figure 5 display the X-ray diffraction topograms at the longitudinal section of the end parts of the crystals that were formed upon the solidification of the final portions of the melt.

A comparative analysis of the grown structures revealed in these fragments showed that the end part of the single crystals obtained under 0.15 MPa pressure occurs of the clear-cut dendritic structure in all volume of the fragment.

It is seen that the crystals remain have a line-type substructure habitual of the more clean matters.

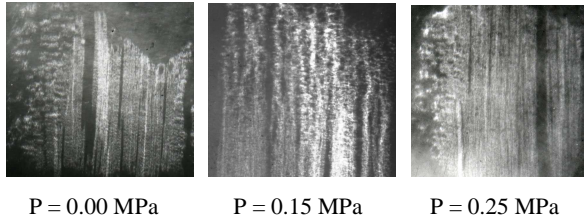

P = 0.00 MPa    P = 0.15 MPa    P = 0.25 MPa

Figure 5. X–ray diffraction topograms of longitudinal polished sections of the end part of the single crystals grown under the various pressures.

## CONCLUSIONS

Upon examination of the dendritic structure of the Al-4 wt % Cu alloy single crystals grown under the action of a gravitational-field component directed along the plane of the crystallization front it has been found that concentration of a copper increases in direction of the field.

Redistribution of the solute component of alloy in the solid solution of the dendrite cells influence on the dispersion of the dendtitic substructure. The most efficient spatial separation of the alloy components is observed under a pressure of 0.25 MPa that is "optimum" in magnitude for the chosen conditions of crystallization.

.